\documentstyle[12pt,graphicx]{article}
\begin{document}
\baselineskip 0.6cm

\def\simgt{\mathrel{\lower2.5pt\vbox{\lineskip=0pt\baselineskip=0pt
           \hbox{$>$}\hbox{$\sim$}}}}
\def\simlt{\mathrel{\lower2.5pt\vbox{\lineskip=0pt\baselineskip=0pt
           \hbox{$<$}\hbox{$\sim$}}}}

\begin{titlepage}

\begin{flushright}
UCB-PTH-05/35 \\
LBNL-59022
\end{flushright}

\vskip 1.5cm

\begin{center}

{\Large \bf 
Natural Little Hierarchy from a Partially Goldstone \\ Twin Higgs}

\vskip 1.0cm

{\large
Z. Chacko$^a$, Yasunori Nomura$^{b,c}$, 
Michele Papucci$^{b,c}$ and Gilad Perez$^c$}

\vskip 0.4cm

$^a$ {\it Department of Physics, University of Arizona,
                Tucson, AZ 85721} \\
$^b$ {\it Department of Physics, University of California,
                Berkeley, CA 94720} \\
$^c$ {\it Theoretical Physics Group, Lawrence Berkeley National Laboratory,
                Berkeley, CA 94720} \\

\vskip 1.2cm

\abstract{We construct a simple theory in which the fine-tuning of the 
standard model is significantly reduced.  Radiative corrections to the 
quadratic part of the scalar potential are constrained to be symmetric 
under a global $U(4) \times U(4)'$ symmetry due to a discrete $Z_2$ 
``twin'' parity, while the quartic part does not possess this symmetry. 
As a consequence, when the global symmetry is broken the Higgs fields
emerge as light pseudo-Goldstone bosons, but with sizable quartic
self-interactions.  This structure allows the cutoff scale, $\Lambda$, 
to be raised to the multi-TeV region without significant fine-tuning. 
In the minimal version of the theory, the amount of fine-tuning is 
about $15\%$ for $\Lambda = 5~{\rm TeV}$, while it is about $30\%$ in 
an extended model.  This provides a solution to the little hierarchy 
problem.  In the minimal model, the ``visible'' particle content is 
exactly that of the two Higgs doublet standard model, while the extended 
model also contains extra vector-like fermions with masses $\approx 
(1\!\sim\!2)~{\rm TeV}$.  At the LHC, our minimal model may appear 
exactly as the two Higgs doublet standard model, and new physics 
responsible for cutting off the divergences of the Higgs mass-squared 
parameter may not be discovered.  Several possible processes that may 
be used to discriminate our model from the simple two Higgs doublet 
model are discussed for the LHC and for a linear collider.}

\end{center}
\end{titlepage}

\section{Introduction}
\label{sec:intro}

Despite its tremendous phenomenological success, the standard model 
is an incomplete theory.  In the standard model, the Higgs mass-squared 
parameter receives radiative corrections of order the cutoff scale 
squared, implying the existence of some new physics at a scale not much 
larger than the scale of electroweak symmetry breaking.  On the other 
hand, experiments have not found any convincing sign of such physics so 
far: the scale suppressing nonrenormalizable operators must be larger 
than several TeV.  This suggests that the new physics must cut off the 
corrections to the Higgs mass-squared parameter without much affecting 
the other sectors of the standard model.  What is this new physics and 
how can we find it?

An interesting idea to control radiative corrections to the Higgs potential 
is to consider it to be the pseudo-Goldstone boson (PGB) of some broken 
global symmetry~\cite{Kaplan:1983fs}.  The actual implementation of this 
idea, however, is not so simple.  The Higgs potential possesses a global 
symmetry at tree level, which is explicitly broken by the electroweak 
gauge and Yukawa interactions.  These explicit breakings then generate the 
potential for the Higgs field $h$ at loop level.  This itself, however, 
does not help much because the generated Higgs mass-squared parameter 
$m_h^2$ is of order $L \Lambda^2$, where $\Lambda$ is the cutoff and $L$ 
is the one-loop factor: $L \approx g^2 C/16\pi^2$ or $y^2 N/16\pi^2$ 
with $g$ and $y$ gauge and Yukawa couplings and $C$ and $N$ 
multiplicity factors.  We have just dropped the tree-level $m_h^2$ 
term in the standard model simply by declaring that the Higgs is a PGB. 
Some progress, however, can be made if we control radiative corrections 
from gauge and Yukawa interactions either by breaking symmetry 
collectively~\cite{Arkani-Hamed:2001nc}, by making the size of the 
gauge group generating the PGB Higgs large and thus separating the 
momentum cutoff scale from the cutoff of the theory~\cite{Contino:2003ve}, 
or by using a discrete symmetry~\cite{Chacko:2005pe}.  In these cases, 
the correction to $m_h^2$ can be cut off below the real cutoff of the 
theory $\Lambda$, so that we can have a perturbative theory describing 
physics above the naive one-loop cutoff scale of the standard model. 
The question for the consistency with the data can then be addressed 
by studying this perturbative physics.

A generic problem for these classes of theories is that since the 
generated Higgs potential is a function of $\cos(h/f)$ and $\sin(h/f)$, 
where $f$ is the decay constant for the PGB Higgs, the vacuum expectation 
value (VEV) of $h$ is naively of order $f$: $\langle h \rangle \approx f$. 
This is not good because the cutoff scale $\Lambda$ is then, at most, of 
order $4\pi f \approx 2~{\rm TeV}$, so that it does not help to understand 
why the deviations from the standard model are experimentally so small. 
There are essentially two ways to evade this problem.  One is to invoke 
a cancellation in the quadratic term in the Higgs potential.  The 
one-loop potential for the PGB Higgs is schematically written as $V(h) 
= L(-\eta_2 f^2 |h|^2 + \eta_4 |h|^4/2 - \eta_6 |h|^6/6 f^2 + \cdots)$, 
where $\eta$'s are naturally of $O(1)$.  Then, if the coefficient $\eta_2$ 
is somehow small, for example due to a cancellation between gauge and 
Yukawa contributions, or if we add an additional term of the form 
$\delta V(h) = \mu^2 |h|^2$ such that $\mu^2 \approx \eta_2 f^2 L$, we 
can obtain $\langle h \rangle \ll f$ and push up the cutoff scale $\Lambda$ 
to be larger than $2~{\rm TeV}$.  The other possibility is to introduce 
an extra quartic coupling, $\delta V(h) = \lambda |h|^4/2$.  This would 
be interesting because we may then obtain $\langle h \rangle \approx 
f L^{1/2} \ll f$ for $\lambda = O(1)$, so that the cutoff may be pushed 
up to $\Lambda \approx \langle h \rangle/L \gg 2~{\rm TeV}$ without 
unnatural cancellations.  The problem is that such a quartic coupling 
also gives a correction to the Higgs mass-squared parameter.  If we set 
the cutoff to be $\Lambda \approx 4\pi f$, the correction is of order 
$\delta m_h^2 \approx (\lambda/16\pi^2)\Lambda^2 \approx \lambda^2 f^2$, 
which is much larger than the corresponding term in the original potential 
$V(h)$.  In order to make this possibility work, therefore, we need some 
mechanism controlling this correction. One such mechanism is collective 
symmetry breaking in little Higgs theories.  Implementing it in realistic 
theories, however, generically requires some non-trivial model building 
efforts~\cite{Arkani-Hamed:2001nc,Gregoire:2002ra}.  Moreover, the 
constraints from the precision electroweak data are often quite 
severe~\cite{Csaki:2002qg}, requiring a further ingredient, such 
as $T$ parity~\cite{Cheng:2003ju}, to make the models fully viable. 

In this paper, we construct a theory which addresses the issues described 
above.  An important ingredient for this construction is the discrete 
$Z_2$ ``twin'' symmetry relating the standard-model fields with their 
mirror partners.  It has recently been shown in~\cite{Chacko:2005pe} 
that this symmetry can be used to control divergences from the gauge 
and Yukawa couplings in PGB Higgs theories.  Using this ``twin Higgs'' 
mechanism, we can construct a simple theory which naturally realizes 
electroweak symmetry breaking.  We show that by introducing an operator 
that explicitly violates the global symmetry but still preserves the 
$Z_2$ symmetry, we can generate an order-one quartic coupling in the 
Higgs potential without giving a quadratically divergent contribution 
to the Higgs mass-squared parameter.  This allows us to push up the 
cutoff scale to the multi-TeV region without significant fine-tuning, 
and thus to solve the little hierarchy problem implied by the mismatch 
between the stability of the electroweak scale and the constraints 
from experiments~\cite{Barbieri:2000gf}.  With an extended top quark 
sector and a mild tuning of order $10\%$, this basic framework allows 
the cutoff scale as high as about $8~{\rm TeV}$.  We assume that our 
theory is weakly coupled at the cutoff scale, although it may be possible 
to extend it to the strongly coupled case.  An interesting aspect of 
the model is that the scalar potential does not possess any approximate 
continuous global symmetry.  The global symmetry is explicitly broken 
by an $O(1)$ amount by a dimensionless quartic coupling.  The gauge 
and Yukawa interactions also break the symmetry by an $O(1)$ amount. 
Nevertheless, the quadratic terms in the scalar potential possess 
an enhanced global symmetry, guaranteed by the discrete $Z_2$ ``twin'' 
symmetry, and this partial global symmetry is sufficient to achieve 
our goals.  The theory has two Higgs doublets, whose couplings to matter 
fields can take either a Type-I, Type-II or mixed form. 

The minimal version of our theory may lead to a potentially embarrassing 
situation at the LHC.  While the theory does not have significant fine-tuning 
in electroweak symmetry breaking, the LHC may just see the (two Higgs doublet) 
standard model, and may not find any new physics responsible for cutting 
off the divergences of the Higgs mass-squared parameter.  This is because 
divergences in the Higgs mass-squared parameter due to the standard 
model fields are canceled by fields that are singlet under the standard 
model gauge group.  The deviations from the simple two Higgs doublet 
model due to these singlet fields can be very small at the LHC. 
The deviations, however, may show up at a linear collider.  This 
demonstrates that it may be too early to give up the concept of 
naturalness even if the LHC does not find any new physics associated 
with the cancellation of the Higgs mass divergences.  Precision Higgs 
studies at a linear collider may be necessary before any firm conclusion 
can be drawn.

In a version of the theory in which the top quark sector is extended and 
the amount of fine-tuning is further reduced, we will find new strongly 
interacting vector-like fermions at the LHC, which are responsible for 
cutting off the radiative correction to the Higgs mass-squared parameter 
from the top quark.  These particles, however, may be the only new 
particle we will find at the LHC beyond the two Higgs doublets, because 
all the other divergences in the standard model can be canceled by fields 
which are singlet under the standard model gauge group.

The organization of the paper is as follows.  In the next section we 
describe the basic structure of our theory.  In section~\ref{sec:rad-corr} 
we calculate radiative corrections to the Higgs potential and show 
that the cutoff scale can be pushed up to the multi-TeV region without 
significant fine-tuning.  In section~\ref{sec:alternative} we extend 
our minimal model so that it allows a smaller fine-tuning and/or 
larger cutoff scale.  We find that the cutoff scale can be raised 
up to about $8~{\rm TeV}$ with a mild tuning of order $10\%$.  In 
section~\ref{sec:strong} we discuss the possibility of making the 
theory strongly coupled at $\Lambda$.  Phenomenology of the model is 
discussed in section~\ref{sec:pheno}, and conclusions are given in 
section~\ref{sec:concl}.

\section{Minimal Theory}
\label{sec:theory}

We consider that our theory is an effective field theory describing 
physics below the cutoff scale $\Lambda$, which is given by specifying 
the Lagrangian at the scale $\Lambda$.  We assume that the theory is 
weakly coupled at $\Lambda$, and that radiative corrections to the 
Higgs mass-squared parameter (at least power divergent ones) are 
cut off at this scale.  We do not need to specify physics above 
$\Lambda$ for the present purpose.  As we will see later, the scale 
$\Lambda$ in our theory can be in the multi-TeV region without 
significant fine-tuning. 

Let us consider two scalar fields $\Phi$ and $\Phi'$ that transform 
as fundamental four-dimensional representations under global $U(4)$ 
and $U(4)'$ symmetries, respectively.  We assume that the tree-level 
potential for $\Phi$ and $\Phi'$ drive non-zero VEVs for $\Phi$ and 
$\Phi'$, breaking $U(4) \rightarrow U(3)$ and $U(4)' \rightarrow U(3)'$, 
respectively.  The $U(4) \times U(4)'$ invariant Lagrangian causing 
such a breaking pattern is%
\footnote{Precisely speaking, the symmetry breaking pattern described 
by Eq.~(\ref{eq:L-Phis}) is $O(8) \times O(8)' \rightarrow O(7) \times 
O(7)'$, but the existence of these larger symmetries does not affect 
any of our argument below.}
\begin{equation}
  {\cal L} = - \eta (|\Phi|^2 - f^2)^2 - \eta' (|\Phi'|^2 - f'^2)^2. 
\label{eq:L-Phis}
\end{equation}
What are the sizes for $\eta$, $\eta'$, $f$ and $f'$?  We take $\eta 
\sim \eta' = O(1)$ because the theory is assumed to be weakly coupled 
at the scale $\Lambda$.  For $f$ and $f'$, we take them to be somewhat 
smaller than the cutoff scale $\Lambda$: $f \sim f' = O(\Lambda/4\pi)$. 
This is crucial to achieve our goal of raising the cutoff, as will 
become clear later.  These values of $f$ and $f'$ are stable under 
radiative corrections, i.e. technically natural.  They may naturally 
arise if $\Phi$ and $\Phi'$ themselves are PGBs of some larger global 
group, say those of $U(5) \times U(5)' \rightarrow U(4) \times U(4)'$, 
but here we simply take $f \sim f' \sim \Lambda/4\pi$ without specifying 
their origin.

We denote the upper and lower halfs of the $\Phi$ ($\Phi'$) field as 
$H_A$ and $H_B$ ($H'_A$ and $H'_B$), respectively.  When $\Phi$ and 
$\Phi'$ develop VEVs
\begin{equation}
  \langle \Phi \rangle 
    = \left\langle \left( \begin{array}{c} H_A \\ 
        \hline H_B \end{array} \right) \right\rangle
    = \left( \begin{array}{c} 0 \\ 0 \\ 
        \hline 0 \\ f \end{array} \right),
\qquad
  \langle \Phi' \rangle 
    = \left\langle \left( \begin{array}{c} H'_A \\ 
        \hline H'_B \end{array} \right) \right\rangle
    = \left( \begin{array}{c} 0 \\ 0 \\ 
        \hline f' \\ 0 \end{array} \right),
\label{eq:Phi-VEVs}
\end{equation}
14~Goldstone bosons appear associated with the breaking $U(4) \times 
U(4)' \rightarrow U(3) \times U(3)'$.  Now, we gauge the $SU(2)_A 
\times U(1)_A \times SU(2)_B \times U(1)_B$ subgroup of $U(4) \times 
U(4)'$.  Here, $SU(2)_A \times U(1)_A$ acts on the upper half components 
of $\Phi$ and $\Phi'$ such that both $H_A$ and $H'_A$ have the quantum 
numbers of ${\bf 2}_{-1/2}$, while $SU(2)_B \times U(1)_B$ on the 
lower half components of $\Phi$ and $\Phi'$ such that $H_B$ and 
$H'_B$ transform as ${\bf 2}_{-1/2}$.  This gauging explicitly breaks 
the $U(4) \times U(4)'$ global symmetry.  Under $SU(2)_A \times U(1)_A$, 
14~Goldstone bosons -- now pseudo-Goldstone bosons (PGBs) -- transform 
as two ${\bf 2}_{-1/2}$'s and six ${\bf 1}_{0}$'s.  We identify 
$SU(2)_A \times U(1)_A$ as $SU(2)_L \times U(1)_Y$ of the standard model. 
We then find that we can obtain two Higgs doublets as PGBs from this 
symmetry breaking pattern.  The stability of the particular form of 
the VEVs in Eq.~(\ref{eq:Phi-VEVs}) will be discussed later.

In what sense are the 14~states PGBs?  Since the theory is weakly coupled 
and the gauging of $SU(2)_A \times U(1)_A \times SU(2)_B \times U(1)_B$ 
explicitly breaks the global $U(4) \times U(4)'$ symmetry by an $O(1)$ 
amount, the theory does not possess an approximate $U(4) \times U(4)'$. 
However, as we will see below, radiative corrections from gauge 
interactions approximately preserve the $U(4) \times U(4)'$ form of the 
scalar potential, if the discrete $Z_2$ symmetry interchanging $SU(2)_A$ 
and $SU(2)_B$, and $U(1)_A$ and $U(1)_B$, is introduced.  In this case, 
$U(4) \times U(4)'$ breaking effects in the scalar potential is of 
order $1/16\pi^2$, and we can still call the 14~states PGBs.

We now impose the $Z_2$ symmetry which interchanges the $A$ and $B$ 
sectors, i.e. $H_A$ and $H_B$, $H'_A$ and $H'_B$, and the gauge 
bosons of $SU(2)_A$ and $SU(2)_B$, and $U(1)_A$ and $U(1)_B$. 
This requires the gauge couplings of $SU(2)_A$ and $SU(2)_B$ to be 
equal, $g_A = g_B = g$, as well as those of $U(1)_A$ and $U(1)_B$, 
$g'_A = g'_B = g'$.  An important consequence of this $Z_2$ 
symmetry is that quadratic divergences from gauge loops to the 
squared-mass parameters for the PGB Higgs bosons are completely 
eliminated~\cite{Chacko:2005pe}.  This is because quadratic divergences 
appear only in the coefficients of the operators quadratic in fields: 
$\delta V = \Lambda^2 (c_A |H_A|^2 + c_B |H_B|^2 + c'_A |H'_A|^2 + 
c'_B |H'_B|^2)$, where $c_A$, $c_B$, $c'_A$ and $c'_B$ are numbers. 
(Operators of the form $H_A^\dagger H'_A + {\rm h.c.}$ and $H_B^\dagger 
H'_B + {\rm h.c.}$ can be forbidden by imposing a $U(1) \times U(1)'$ 
global symmetry; see discussion later.)  Since the $Z_2$ symmetry 
always guarantees that $c_A$ and $c_B$, and $c'_A$ and $c'_B$, are 
equal, quadratically divergent radiative corrections necessarily take 
the $U(4) \times U(4)'$ invariant form: $\delta V = c_A \Lambda^2 
(|H_A|^2 + |H_B|^2) + c'_A \Lambda^2 (|H'_A|^2 + |H'_B|^2)= c_A \Lambda^2 
|\Phi|^2 + c'_A \Lambda^2 |\Phi'|^2$, which do not give any potential 
for the PGBs.  In fact, one can explicitly check in a non-linear sigma 
model that if the gauge couplings of $SU(2)_A$ and $SU(2)_B$, and 
$U(1)_A$ and $U(1)_B$, are the same as dictated by the $Z_2$ symmetry, 
quadratically divergent contributions to the PGB potential are absent. 
The potential for the PGB Higgs arises from operators of the form
\begin{equation}
  \delta V = \xi (|H_A|^4 + |H_B|^4) 
    + \xi' (|H'_A|^4 + |H'_B|^4) + \cdots,
\label{eq:PGB-corr}
\end{equation}
which are $Z_2$ invariant but not $U(4) \times U'(4)$ invariant. 
It is then clear from dimensional analysis that the potential 
for the PGBs is at most logarithmically divergent.%
\footnote{Radiative corrections involving higher dimension operators 
can, of course, generate power divergent corrections to the PGB 
potential, but they are sufficiently small if the theory is weakly 
coupled at $\Lambda$.}
The sizes of the coefficients $\xi$ and $\xi'$ in Eq.~(\ref{eq:PGB-corr}) 
are of order $(g^2/16\pi^2)\ln(\Lambda/f)$, so that the PGB Higgses 
receive squared masses only of order $(g^2 f^2/16\pi^2)\ln(\Lambda/f)$. 

The situation for the Yukawa interactions is similar.  If we make 
the Yukawa couplings $Z_2$ invariant by introducing mirror quarks, 
quadratically divergent radiative corrections to the squared masses 
for the PGB Higgses are eliminated.  For example, for the top quarks 
we introduce mirror quarks $\hat{q}$ and $\hat{\bar{u}}$, which are 
singlet under $SU(2)_L \times U(1)_Y$ and transform as ${\bf 2}_{1/6}$ 
and ${\bf 1}_{-2/3}$ under $SU(2)_B \times U(1)_B$, in addition to our 
quarks $q$ and $\bar{u}$, which transform as ${\bf 2}_{1/6}$ and 
${\bf 1}_{-2/3}$ under $SU(2)_L \times U(1)_Y$ and are singlet under 
$SU(2)_B \times U(1)_B$.  For color interactions, we assume that our 
quarks and mirror quarks are charged under $SU(3)_A$ and $SU(3)_B$ 
gauge interactions, respectively, where $SU(3)_A$ is identified as 
the standard model color group: $SU(3)_A \equiv SU(3)_C$.  Writing 
the $Z_2$-invariant Yukawa coupling 
\begin{equation}
  {\cal L}_{\rm top} 
    = y_t (q \bar{u} H_A^\dagger + \hat{q} \hat{\bar{u}} H_B^\dagger),
\label{eq:top-Yukawa}
\end{equation}
the PGBs do not receive any quadratically divergent contributions 
from this coupling.  Here, we couple only $\Phi = (H_A | H_B)$ to the 
top quarks, and not $\Phi' = (H'_A | H'_B)$.  Such a situation can 
be easily arranged, for example, by considering that the $U(1) \times 
U(1)'$ subgroup of the $U(4) \times U(4)'$ global symmetry is an exact 
(anomalous) global symmetry and assigning appropriate charges to the 
quark fields.  The symmetry $U(1) \times U(1)'$ will also be discussed 
later when we introduce an explicit $U(4) \times U(4)'$ breaking 
operator in the scalar potential.

There are two ways to introduce the bottom Yukawa coupling into the 
theory, without introducing dangerous flavor changing neutral currents. 
One way is to couple only $\Phi = (H_A | H_B)$ to the bottom quarks:
\begin{equation}
  {\cal L}_{\rm bottom} 
    = y_b (q \bar{d} H_A + \hat{q} \hat{\bar{d}} H_B),
\label{eq:bottom-Yukawa-1}
\end{equation}
where $\bar{d}$ is the right-handed bottom quark transforming as 
${\bf 1}_{1/3}$ under $SU(2)_L \times U(1)_Y$ and singlet under 
$SU(2)_B \times U(1)_B$, while $\hat{\bar{d}}$ is its mirror partner 
transforming as ${\bf 1}_{1/3}$ under $SU(2)_B \times U(1)_B$ and 
singlet under $SU(2)_L \times U(1)_L$.  The other way is to couple 
only $\Phi' = (H'_A | H'_B)$ to the bottom quarks:
\begin{equation}
  {\cal L}_{\rm bottom} 
    = y_b (q \bar{d} H'_A + \hat{q} \hat{\bar{d}} H'_B).
\label{eq:bottom-Yukawa-2}
\end{equation}
Since our two PGB-Higgs doublets essentially come from $H_A$ 
and $H'_A$, the two cases of Eqs.~(\ref{eq:bottom-Yukawa-1}) and 
(\ref{eq:bottom-Yukawa-2}) lead, respectively, to Type-I and Type-II 
Higgs doublet theories.  The Yukawa couplings for lighter quarks can 
be obtained by making $y_t$ and $y_b$ to $3 \times 3$ matrices.  The 
Yukawa couplings for leptons are introduced analogously to the down-type 
quarks, but the choice between Eqs.~(\ref{eq:bottom-Yukawa-1}) and 
(\ref{eq:bottom-Yukawa-2}) can be made independently from that for 
the down-type quarks.  The particular form of the couplings in 
Eq.~(\ref{eq:bottom-Yukawa-1}) or (\ref{eq:bottom-Yukawa-2}) can, 
again, be ensured by considering that the $U(1) \times U(1)'$ subgroup 
of $U(4) \times U(4)'$ is exact and by assigning appropriate $U(1) 
\times U(1)'$ charges to $q$, $\bar{d}$, $\hat{q}$ and $\hat{\bar{d}}$ 
(and to the corresponding lepton fields). 

With these structures for gauge and Yukawa interactions, radiative 
corrections to the squared masses for the PGB Higgses can be made small 
to the level of $O((f^2/16\pi^2)\ln(\Lambda/f))$.  This itself, however, 
does not achieve our goal of naturally raising the cutoff $\Lambda$ 
to the multi-TeV region. Since our Higgs fields, $h$, are PGBs, their 
potential generated by gauge and Yukawa interactions is a function 
of $\cos(h/f)$ and $\sin(h/f)$, giving $\langle h \rangle \approx 
f \approx 200~{\rm GeV}$.  This in turn implies $\Lambda \simlt 
4\pi f \approx 2~{\rm TeV}$.  The source of the problem is that 
while the Higgs mass-squared parameters are suppressed to the level 
of $O((f^2/16\pi^2)\ln(\Lambda/f))$, the quartic couplings are also 
suppressed and of order $O((1/16\pi^2)\ln(\Lambda/f))$.  Moreover, the 
stability of the particular form of the VEVs in Eq.~(\ref{eq:Phi-VEVs}) 
is not obvious at this stage, without a detailed study of the PGB 
potential generated at loop level.

We now present a mechanism addressing these issues and present 
a realistic theory in which $\Lambda$ can be raised to the multi-TeV 
region without a significant fine-tuning.  Suppose we introduce 
a tree-level operator
\begin{equation}
  {\cal L}_H = - \lambda (|H_A^\dagger H'_A|^2 + |H_B^\dagger H'_B|^2),
\label{eq:quartic}
\end{equation}
which explicitly violates the global $U(4) \times U(4)'$ symmetry 
but preserves the $Z_2$ symmetry.  We take the coupling $\lambda$ to 
be of $O(1)$.  We then find that the operator of Eq.~(\ref{eq:quartic}) 
gives an order-one quartic coupling for the two PGB-Higgs doublets 
without giving large squared masses, and at the same time stabilizes 
the desired vacuum of Eq.~(\ref{eq:Phi-VEVs}).  To see this explicitly, 
we expand the $\Phi$ and $\Phi'$ fields as
\begin{equation}
  \Phi = \left( \begin{array}{c} H_A \\ \hline H_B \end{array} \right)
  = \exp\left[ \frac{i}{f} \left( \begin{array}{cccc}
    0 & 0 & 0 & h_1 \\ 0 & 0 & 0 & h_2 \\ 0 & 0 & 0 & \frac{a+ib}{\sqrt{2}} \\
    h_1^\dagger & h_2^\dagger & \frac{a-ib}{\sqrt{2}} & \frac{c}{\sqrt{2}} 
    \end{array} \right) \right] 
    \left( \begin{array}{c} 0 \\ 0 \\ 0 \\ f \end{array} \right),
\label{eq:nonlinear-Phi}
\end{equation}
and
\begin{equation}
  \Phi' = \left( \begin{array}{c} H'_A \\ \hline H'_B \end{array} \right)
  = \exp\left[ \frac{i}{f'} \left( \begin{array}{cccc}
    0 & 0 & h'_1 & 0 \\ 0 & 0 & h'_2 & 0 \\ 
    h_1^{\prime\dagger} & h_2^{\prime\dagger} & 
    \frac{c'}{\sqrt{2}} & \frac{a'-ib'}{\sqrt{2}} \\
    0 & 0 & \frac{a'+ib'}{\sqrt{2}} & 0
    \end{array} \right) \right] 
    \left( \begin{array}{c} 0 \\ 0 \\ f' \\ 0 \end{array} \right),
\label{eq:nonlinear-Phi'}
\end{equation}
respectively, where $h=(h_1, h_2)$ and $h'=(h'_1, h'_2)$ are the two 
PGB Higgs doublets, and $a$, $b$, $c$, $a'$, $b'$ and $c'$ are the six 
singlet PGBs, of the spontaneous $U(4) \times U(4)' \rightarrow U(3) 
\times U(3)'$ breaking.  Here the PGB fields are canonically normalized, 
and we have neglected the radial excitation modes of $\Phi$ and $\Phi'$. 
Substituting Eqs.~(\ref{eq:nonlinear-Phi},~\ref{eq:nonlinear-Phi'}) 
into Eq.~(\ref{eq:quartic}), we obtain
\begin{equation}
  V = -{\cal L}_H = \lambda |h^\dagger h'|^2 
    + \frac{1}{2} \lambda (f^2 + f'^2) (\tilde{a}^2 + \tilde{b}^2) 
    + \cdots,
\label{eq:explicit}
\end{equation}
where $\tilde{a} \equiv (f'a-fa')/\sqrt{f^2+f'^2}$ and $\tilde{b} \equiv 
(f'b+fb')/\sqrt{f^2+f'^2}$ are canonically normalized singlet PGBs that 
are not eaten by the massive $SU(2)_B \times U(1)_B$ gauge bosons.  (The 
eaten modes are $(fa+f'a')/\sqrt{f^2+f'^2}$, $(fb-f'b')/\sqrt{f^2+f'^2}$, 
$c$ and $c'$.)  This explicitly shows that for $\lambda > 0$ the 
operator of Eq.~(\ref{eq:quartic}) gives a quartic coupling $\lambda$ 
to the two PGB-Higgs doublets, $h$ and $h'$, and the vacuum of 
Eq.~(\ref{eq:Phi-VEVs}) is stabilized by the masses of $\tilde{a}$ and 
$\tilde{b}$, $m_{\tilde{a}}^2 = m_{\tilde{b}}^2 = \lambda(f^2+f'^2) > 0$. 
A similar operator has also been used in little Higgs theories to obtain 
a tree-level quartic coupling~\cite{Kaplan:2003uc}.

An interesting point here is that while the operator of 
Eq.~(\ref{eq:quartic}) introduces an order-one explicit breaking 
of the global $U(4) \times U(4)'$ symmetry to the scalar potential (and 
hence the $O(f)$ masses for $\tilde{a}$ and $\tilde{b}$), the Higgs 
doublets do not obtain masses of order $f$.  The masses are generated 
at loop level, but because of $Z_2$ invariance they are generated only 
through {\it quartic} couplings between $H$'s, such as the ones in 
Eq.~(\ref{eq:PGB-corr}).  The coefficients of these operators, e.g. 
$\xi$ and $\xi'$ in Eq.~(\ref{eq:PGB-corr}), are at most of order 
$(1/16\pi^2)\ln(\Lambda/f)$, since they are generated at loop level 
and the theory is weakly coupled.%
\footnote{The argument here shows that the theory could potentially have 
a problem if it is strongly coupled e.g. $\eta \sim \eta' \sim 4\pi$, 
because then the coefficients $\xi$ and $\xi'$ may receive corrections 
of order e.g. $\eta \lambda/16\pi^2, \eta g^2/16\pi^2 \sim 1$, giving 
the Higgs squared masses of order $f^2$, which would be too large for 
our purpose~\cite{Barbieri:2005ri}.  We will discuss this issue in 
section~\ref{sec:strong}.}
This guarantees that radiatively generated Higgs squared masses cannot 
be larger than of order $(f^2/16\pi^2)\ln(\Lambda/f)$.  We note here 
that our 14~states are no longer ``PGBs'' in the usual sense, since the 
potential as a whole does not possess an approximate $U(4) \times U(4)'$ 
symmetry: it is broken by an $O(1)$ amount by $\lambda$.  What ensures 
the stability of the potential here under radiative corrections from 
explicit symmetry breaking interactions is the  ``partial $U(4) \times 
U(4)'$ symmetry'' --- $U(4) \times U(4)'$ possessed only by the 
quadratic terms of the scalar potential, which arises as a consequence 
of the discrete $Z_2$ symmetry of the theory.

It is technically natural to introduce only the operator of 
Eq.~(\ref{eq:quartic}) as an $O(1)$ $U(4) \times U(4)'$-violating 
effect in the scalar potential.  Other $U(4) \times U(4)'$-violating 
terms are generated at loop level, but they are at most of order 
$1/16\pi^2$.  In fact, this particular explicit symmetry breaking 
pattern may be justified by assuming certain structure for the ultraviolet 
theory above $\Lambda$.  Imagine, for example, that the operator of 
Eq.~(\ref{eq:quartic}) is generated by tree-level exchanges of auxiliary 
scalar fields that have $Z_2 \times U(1) \times U(1)'$ invariant 
trilinear couplings between primed, non-primed and the auxiliary 
fields.  Then, the only $U(4) \times U(4)'$-violating operators generated 
at tree level are the one in Eq.~(\ref{eq:quartic}) and an operator 
$H_A^\dagger H'_A H_B H_B^{\prime\dagger} + {\rm h.c.}$  We find that 
the existence of the latter operator with an $O(1)$ coefficient does 
not change any of the basic aspects of the model.  This operator, however, 
can also be forbidden if we impose a discrete $Z_2$ ``chiral'' symmetry: 
$(H_A | H_B) \leftrightarrow (H_A | H_B)$ and $(H'_A | H'_B) 
\leftrightarrow (H'_A | -\!H'_B)$.  Below we impose this $Z_2$ symmetry 
and set the coefficient of the above operator to be zero for simplicity. 
We also impose the $U(1) \times U(1)' \subset U(4) \times U(4)'$ 
symmetry as an exact (anomalous) global symmetry of the model.  This 
suppresses the operator $\Phi^\dagger \Phi' + {\rm h.c.}$, whose 
coefficient must be of order $f^2$ or smaller since otherwise some 
of the modes needed to cancel quadratic divergences to the Higgs 
squared masses become too heavy. 

Summarizing so far, the Lagrangian of our theory is given by the scalar 
potential of Eqs.~(\ref{eq:L-Phis},~\ref{eq:quartic}) and the Yukawa 
couplings of either Eqs.~(\ref{eq:top-Yukawa},~\ref{eq:bottom-Yukawa-1}) 
or Eqs.~(\ref{eq:top-Yukawa},~\ref{eq:bottom-Yukawa-2}).  At scales 
below $f$, the theory contains the standard model quarks and leptons 
as well as the two Higgs doublets, $h$ and $h'$, which have the 
following dimensionless couplings:
\begin{equation}
  {\cal L} = y_t q \bar{u} h^\dagger + y_b q \bar{d} h^{(\prime)} 
    + y_\tau l \bar{e} h^{(\prime)} - V(h,h'),
\label{eq:type-I}
\end{equation}
where $l$ and $\bar{e}$ are the doublet and singlet lepton fields, 
respectively, and the Higgs potential $V(h,h')$ contains the 
tree-level quartic coupling $\lambda |h^\dagger h'|^2$ as well 
as radiatively generated Higgs mass-squared parameters of order 
$(f^2/16\pi^2)\ln(\Lambda/f)$.  The Higgs field to which the down-type 
quarks and charged leptons couple can be either $h$ or $h'$, depending 
on their $U(1) \times U(1)'$ charges, and the choice can be made 
independently for the down-type quarks and charged leptons (if there 
is no quark-lepton unification in the fundamental theory).

Because of the particular form of the Higgs quartic coupling arising 
from the operator of Eq.~(\ref{eq:quartic}), $\lambda |h^\dagger h'|^2$, 
the squared mass parameters for $h$ and $h'$ must both be positive 
to ensure the absence of a runaway direction in the potential. 
Electroweak symmetry breaking then must be caused by the term 
$h^\dagger h' + {\rm h.c.}$, by making one of the eigenvalues in 
the Higgs mass-squared matrix negative.  We assume that these mass 
terms are generated by soft $Z_2$-symmetry breaking operators
\begin{equation}
  {\cal L}_{\rm soft} = - \mu^2 |H_A|^2 - \mu'^2 |H'_A|^2 
    + (b H_A^\dagger H'_A + {\rm h.c.}),
\label{eq:Z2-breaking}
\end{equation}
where we take parameters $\mu^2$, $\mu'^2$ and $b$ to be of order 
$(f^2/16\pi^2)\ln(\Lambda/f)$, which is technically natural.  The 
Higgs potential $V(h,h')$ is then given by
\begin{equation}
  V(h,h') = m^2 |h|^2 + m'^2 |h'|^2 - (b\, h^\dagger h' + {\rm h.c.}) 
            + \lambda |h^\dagger h'|^2,
\label{eq:Higgs-pot}
\end{equation}
where $m^2$ and $m'^2$ are given at tree level by $\mu^2$ and $\mu'^2$, 
respectively, but they also receive radiative corrections of order 
$(f^2/16\pi^2)\ln(\Lambda/f)$.  Here, we have suppressed radiatively 
generated quartic terms as well as higher order terms.  The conditions 
for having the stable minimum breaking the electroweak symmetry are 
\begin{equation}
  m^2 > 0, \qquad m'^2 > 0, \qquad |b|^2 > m^2 m'^2.
\label{eq:ewsb-cond}
\end{equation}
With these conditions satisfied, we expect to obtain the desired hierarchy
\begin{equation}
  \Lambda \approx 4\pi f \approx (4\pi)^2 v,
\label{eq:hierarchy}
\end{equation}
without a significant fine-tuning, where $v \equiv (\langle h \rangle^2 
+ \langle h' \rangle^2)^{1/2} \simeq 174~{\rm GeV}$.  To reliably estimate 
how large we can make $\Lambda$ without fine-tuning, however, we need 
to calculate radiative correction to $m^2$ and $m'^2$ from top-Yukawa, 
gauge and Higgs-quartic interactions, and carefully study fine-tuning 
required to obtain successful electroweak symmetry breaking.  This will 
be performed in the next section, where we find that the estimate of 
Eq.~(\ref{eq:hierarchy}) is somewhat too optimistic.

\section{Analysis of Fine-Tuning}
\label{sec:rad-corr}

Since radiative corrections to the Higgs mass squared parameters 
in our theory come only from the {\it quartic} terms in the scalar 
potential, we can reliably estimate their sizes at the leading-log 
level.  Specifically, given the Lagrangian of Eqs.~(\ref{eq:L-Phis},%
~\ref{eq:top-Yukawa},~\ref{eq:quartic}), we can evaluate the 
coefficients of $U(4) \times U(4)'$-violating operators
\begin{eqnarray}
  {\cal L} &=& - \xi (|H_A|^4 + |H_B|^4) - \xi' (|H'_A|^4 + |H'_B|^4)
  - \kappa (|H_A|^2 |H'_A|^2 + |H_B|^2 |H'_B|^2),
\label{eq:violating}
\end{eqnarray}
that give masses for the Higgs doublets, where we have kept only operators 
that preserve $U(1) \times U(1)'$ and the ``chiral'' $Z_2$ symmetry. 
We can then obtain expressions for radiative corrections to the Higgs 
mass-squared parameters in terms of the renormalized $f$ and $f'$ 
parameters.  This determines how large we can make $f$ and $f'$ without 
severe fine-tuning, which in turn determines how large the cutoff scale 
$\Lambda$ can be.  In our analysis we assume either that down-type quarks 
and leptons couple to $h$ or that the ratio $\langle h \rangle/\langle 
h' \rangle$ is not very large, so that only the relevant Yukawa coupling 
is the top Yukawa coupling.  An extension to include the bottom and tau 
Yukawa couplings, however, is straightforward.

At the one-loop leading-log level, the coefficients $\xi$, $\xi'$ and 
$\kappa$ in Eq.~(\ref{eq:violating}) receive the following radiative 
corrections:
\begin{eqnarray}
  \delta \xi &=&
    \frac{1}{16\pi^2} \biggl( 6 y_t^4 
      - \frac{9}{8}g^4 - \frac{3}{4}g^2g'^2 - \frac{3}{8}g'^4   
      - \lambda^2 \biggr) \ln\frac{\Lambda}{f},
\label{eq:del-xi} \\
  \delta \xi' &=&
    \frac{1}{16\pi^2} \biggl( 
      - \frac{9}{8}g^4 - \frac{3}{4}g^2g'^2 - \frac{3}{8}g'^4   
      -\lambda^2 \biggr) \ln\frac{\Lambda}{f},
\label{eq:del-xi'} \\
  \delta \kappa &=&
    \frac{1}{16\pi^2} \biggl( 
      - \frac{9}{4}g^4 + \frac{3}{2}g^2g'^2 - \frac{3}{4}g'^4 
      - 2 \lambda^2 \biggr) \ln\frac{\Lambda}{f},
\label{eq:del-kap}
\end{eqnarray}
where $y_t$ is the top Yukawa coupling in Eq.~(\ref{eq:top-Yukawa}), 
$\eta$, $\eta'$ and $\lambda$ are couplings in Eqs.~(\ref{eq:L-Phis},%
~\ref{eq:quartic}), $g$ is the $Z_2$ invariant gauge coupling of 
$SU(2)_A \equiv SU(2)_L$ and $SU(2)_B$, and $g'$ that of $U(1)_A 
\equiv U(1)_Y$ and $U(1)_B$.  The finite pieces depend on the unknown 
ultraviolet theory and do not have a real physical meaning in the 
effective theory.  From these equations, we obtain the expressions for 
the corrections to the Higgs mass-squared parameters $m^2$, $m'^2$ and 
$b$ in Eq.~(\ref{eq:Higgs-pot}):
\begin{eqnarray}
  \delta m^2 &=&
    -2 f^2\, \delta\xi - f'^2\, \delta\kappa,
\label{eq:del-m2} \\
  \delta m'^2 &=&
    -2 f'^2\, \delta\xi' - f^2\, \delta\kappa,
\label{eq:del-m2'} \\
  \delta b &=& 0,
\label{eq:del-b}
\end{eqnarray}
which are of order $(f^2/16\pi^2)\ln(\Lambda/f)$.  Contributions 
arising from renormalizations of the $\mu^2$ and $\mu'^2$ parameters in 
Eq.~(\ref{eq:Z2-breaking}) are of order $(f^2/(16\pi^2)^2)\ln(\Lambda/f)$ 
and thus negligible. 

What is the amount of fine-tuning for this potential?  Let us first 
see that the fine-tuning parameter $\Delta^{-1}$ can be represented in 
terms of the Lagrangian parameters and/or physical Higgs boson masses in 
the following way~\cite{Nomura:2005rk}.  The equations determining the 
minimum of the potential, Eq.~(\ref{eq:Higgs-pot}), can be written as 
\begin{eqnarray}
  && \tan^2\!\beta = \frac{m'^2}{m^2},
\label{eq:min-1}
\\
  && \lambda v^2 = \frac{2 b}{\sin 2 \beta} - (m^2 + m'^2),
\label{eq:min-2}
\end{eqnarray}
where $\tan\beta \equiv \langle h \rangle/\langle h' \rangle$ and 
$v \equiv (\langle h \rangle^2 + \langle h' \rangle^2)^{1/2} \simeq 
174~{\rm GeV}$ is the electroweak scale.  We then find that the only 
source of a potential unnatural cancellation is in the right-hand-side 
of Eq.~(\ref{eq:min-2}), and that the fine-tuning parameter $\Delta^{-1}$ 
is approximately given by the ratio of $\lambda v^2$ and $m^2+m'^2$ 
(or $2 b/\sin2\beta$): $\Delta^{-1} \approx \lambda v^2/(m^2+m'^2)$. 
(Note that $m^2$ and $m'^2$ are both positive, so that they cannot 
be canceled with each other.)  On the other hand, the masses of the 
physical Higgs bosons are given by
\begin{eqnarray}
  && m_{A^0}^2 = m^2 + m'^2 + \lambda v^2,
\label{eq:mA} \\
  && m_{H^\pm}^2 = m^2 + m'^2,
\label{eq:mHpm} \\
  && m_{H^0,h^0}^2 = \frac{1}{2} \Bigl\{ m_{A^0}^2 \pm \sqrt{m_{A^0}^4 
    \cos^2\!2\beta+(m_{A^0}^2-2\lambda v^2)^2 \sin^2\!2\beta} 
  \Bigr\},
\label{eq:mH-mh}
\end{eqnarray}
where $A^0$, $H^\pm$, $H^0$, and $h^0$ represent the pseudoscalar, 
charged, heavier neutral, and lighter neutral Higgs bosons, respectively. 
Assuming that the lighter neutral Higgs boson $h^0$ is somewhat lighter 
than the other Higgs bosons, we obtain
\begin{eqnarray}
  && m_{H^0}^2 \simeq m_{A^0}^2 = m^2 + m'^2 + \lambda v^2,
\label{eq:app-mH-mA}\\
  && m_{H^\pm}^2 = m^2 + m'^2,
\label{eq:app-mHpm}\\
  && m_{h^0}^2 \simeq \lambda v^2 \sin^2\!2\beta.
\label{eq:app-mh}
\end{eqnarray}
The fine-tuning parameter can then be written as 
\begin{equation}
  \Delta^{-1} \approx \frac{\lambda v^2}{m^2+m'^2}
    \simeq \frac{m_{h^0}^2}{m_{H^\pm}^2 \sin^2\!2\beta}.
\label{eq:ft-para}
\end{equation}
For $\tan\beta$ not much larger than $1$, e.g. $\tan\beta \simlt 2$, 
this simplifies further to $\Delta^{-1} \sim m_{h^0}^2/m_{H^\pm}^2$.

We now estimate how high we can push up the cutoff scale $\Lambda$. 
Here we assume $f \simeq f'$ for simplicity.  First, we rewrite 
Eq.~(\ref{eq:ft-para}) using Eqs.~(\ref{eq:min-1},~\ref{eq:app-mh}) 
as $\Delta^{-1} \simeq \lambda v^2/((1+\tan^2\!\beta) m^2) \simeq 
m_{h^0}^2/(4 m^2 \sin^2\!\beta)$.  The parameter $m^2$ receives 
contributions both at tree level, $m^2|_{\rm tree} = \mu^2$, and at 
radiative level, $\delta m^2$ in Eq.~(\ref{eq:del-m2}).  In order to 
avoid unnatural cancellations among these contributions, $m^2$ itself 
must be at least of the same size as the largest radiative contribution. 
For $f \simeq f'$, the largest one comes either from the top loop 
contribution:
\begin{equation}
  \delta m^2|_{\rm top} = -\frac{3y_t^4}{4\pi^2} f^2 \ln\frac{\Lambda}{f},
\label{eq:top-contr}
\end{equation}
where we have used Eqs.~(\ref{eq:del-xi},~\ref{eq:del-m2}), or from 
the Higgs quartic contribution:
\begin{equation}
  \delta m^2|_{H^4} 
    = \frac{\lambda^2}{8\pi^2} (f^2+f'^2) \ln\frac{\Lambda}{f},
\label{eq:Higgs-contr}
\end{equation}
where we have used Eqs.~(\ref{eq:del-xi},~\ref{eq:del-kap},~\ref{eq:del-m2}). 
Now, setting $m^2 \approx |\delta m^2|_{\rm top}|$ and using $m_t = y_t v 
\sin\beta$, the contribution to the fine-tuning parameter from top loop 
can be written as:
\begin{equation}
  \Delta^{-1}|_{\rm top} \equiv 
    \frac{m_{h^0}^2}{4 |\delta m^2|_{\rm top}| \sin^2\!\beta} \approx 
    \frac{\pi^2 v^4 m_{h^0}^2 \sin^2\!\beta}{3 m_t^4 f^2 \ln(\Lambda/f)}
  \simeq 2 \frac{m_{h^0}^2 \sin^2\!\beta}{f^2},
\label{eq:fine-tuning-top}
\end{equation}
where we have used $m_t = m_t|_{\rm pole}(1+g_3^2/3\pi^2)^{-1} \simeq 
166~{\rm GeV}$ and $\ln(\Lambda/f) \simeq \ln(2\pi)$ in the last 
equation (see below).  The contribution from quartic loop, on the 
other hand, can be written using Eq.~(\ref{eq:app-mh}) as
\begin{equation}
  \Delta^{-1}|_{H^4} \equiv 
    \frac{m_{h^0}^2}{4 \delta m^2|_{H^4} \sin^2\!\beta} \approx 
    \frac{32 \pi^2 v^4 \sin^2\!\beta \cos^4\!\beta}
      {m_{h^0}^2 (f^2 + f'^2) \ln(\Lambda/f)}
  \simeq \frac{\sin^2\!\beta \cos^4\!\beta}{m_{h^0}^2 f^2}(530~{\rm GeV})^4,
\label{eq:fine-tuning-Higgs}
\end{equation}
where we have set $f=f'$ in the last equation.  The fine-tuning parameter 
$\Delta^{-1}$ is then given by
\begin{equation}
  \Delta^{-1} = {\rm min} \Bigl\{ \Delta^{-1}|_{\rm top},\,\, 
    \Delta^{-1}|_{H^4} \Bigr\}.
\label{eq:fine-tuning}
\end{equation}

\ From Eqs.~(\ref{eq:fine-tuning-top},~\ref{eq:fine-tuning-Higgs},%
~\ref{eq:fine-tuning}), we find that a maximum value for $\Delta^{-1}$ is 
obtained for $m_{h^0}^2 \simeq (530~{\rm GeV})^2 \cos^2\!\beta\, /\sqrt{2}$, 
with the value $\Delta^{-1} \simeq (320~{\rm GeV}/f)^2 \sin^2\!2\beta$. 
Under the constraint from precision electroweak measurements, $m_{h^0} 
\simlt 250~{\rm GeV}$~\cite{unknown:2005em}, this occurs when $m_{h^0} 
\simeq 250~{\rm GeV}$ and $\tan\beta \simeq 1.5$, and the largest 
value of $f$ for a fixed $\Delta^{-1}$ is given by
\begin{equation}
  f_{\rm max} \approx 650~{\rm GeV} 
    \left( \frac{20\%}{\Delta^{-1}} \right)^{1/2}.
\label{eq:fmax}
\end{equation}
(This value can also be reproduced by taking a complete one-loop 
effective potential into account and seeing the sensitivity of $v^2$ with 
respect to the parameter $b$.)  The relation between $\Lambda$ and $f$ 
is not calculable because $f^2$ receives quadratically divergent radiative 
corrections proportional to $\Lambda^2$.  The relation, however, can be 
estimated using a naive scaling argument:
\begin{equation}
  f^2 \approx \frac{N_f}{16\pi^2} \Lambda^2,
\label{eq:f-Lambda}
\end{equation}
where $N_f$ is the number of ``flavors'', which is $4$ in our case. 
For $\eta \simgt 1$, this relation roughly agrees with the result 
obtained naively by calculating the coefficient of the quadratic 
divergence of $f^2$ in the effective theory.  For smaller $\eta$, 
the hierarchy between $\Lambda$ and $f$ may be smaller because of the 
top Yukawa contribution to the $\Phi$ mass term.  From Eqs.~(\ref{eq:fmax},%
~\ref{eq:f-Lambda}) we obtain the maximum value of the cutoff:
\begin{equation}
  \Lambda_{\rm max} \approx 4~{\rm TeV} 
    \left( \frac{20\%}{\Delta^{-1}} \right)^{1/2}.
\label{eq:Lmax}
\end{equation}
To evade the experimental constraints from higher dimension operators 
we need to have $\Lambda \simgt 5~{\rm TeV}$.%
\footnote{Some higher dimension operators, e.g. $h^\dagger \sigma^a h 
W^a_{\mu\nu} B^{\mu\nu}$, require $\Lambda \simeq 10~{\rm TeV}$ if the 
coefficients are really $1$.  They are, however, expected to carry factors 
of, e.g., $gg'$ in front, in which case the bound on $\Lambda$ is somewhat 
weaker and of order several TeV.}
Our theory requires (only) a mild fine-tuning of about 
\begin{equation}
  \Delta^{-1} \approx 14\% \left( \frac{5~{\rm TeV}}{\Lambda} \right)^2,
\label{eq:ft-model}
\end{equation}
to achieve this.  If we restrict ourselves to $m_{h^0} \simlt 200~{\rm 
GeV}$, this number becomes $\approx 10\%$.  We note here that the 
precise number in Eq.~(\ref{eq:ft-model}) is subject to uncertainties 
of order $(20\!\sim\!30)\%$, for example, due to finite corrections at 
$\Lambda$ to $\delta m^2$, $\delta m'^2$ and $f^2$.

We find that our theory does not really give the naive hierarchies of 
Eq.~(\ref{eq:hierarchy}).  This is because for large $\tan\beta$, we 
need to have $m'^2$ much larger than $m^2$ (see Eq.~(\ref{eq:min-1})), 
so that we need to cancel this large $m'^2$ with the $b$ term in the 
minimization equation of Eq.~(\ref{eq:min-2}).  The Higgs quartic 
coupling $\lambda$ also becomes large in this region, and fine-tuning 
from this parameter, $\Delta^{-1}|_{H^4}$, also becomes severe.  For 
smaller $\tan\beta$, on the other hand, the top Yukawa coupling becomes 
larger, making fine-tuning from top loop, $\Delta^{-1}|_{\rm top}$, 
worse.  This is especially the case because the top radiative correction 
to $m^2$ is proportional to $y_t^4$ (see Eq.~(\ref{eq:top-contr})). 
Here the extra $y_t^2$ in addition to the naive $y_t^2$ comes from the 
fact that the particle cutting off the top divergence in the standard 
model is the mirror top quark, whose mass is proportional to $y_t$: 
$m_{\hat{q}} = y_t f$.  In the next section we present a theory in 
which the contribution from the standard model top quark is canceled 
by the $U(4)$ partner of the top quark, in which case the logarithmic 
sensitivity of the top contribution to $\Lambda$ is eliminated and 
we can achieve further reduction of fine-tuning (or push up $\Lambda$ 
further for a given $\Delta^{-1}$). 

To assess the degree of success here, let us compare our theory with 
the standard model (with the tree-level Higgs mass-squared parameter 
set to zero by hand).  In the standard model, the Higgs mass-squared 
parameter receives quadratically divergent contribution, whose cutoff will 
in general be different from that of $f^2$ in Eq.~(\ref{eq:f-Lambda}). 
It is, therefore, not possible to make a real comparison between the 
two theories.  Nevertheless, if we naively take the quadratic divergent 
part from the top loop, $\delta m_h^2 = -(3 y_t^2/8\pi^2) \Lambda^2$, 
and simply define the fine-tuning parameter for the standard model by 
$\Delta^{-1}_{\rm SM} = \lambda v^2/|\delta m_h^2|$, the standard model 
gives
\begin{equation}
  \Delta^{-1}_{\rm SM} 
    \approx 3.5\% \left( \frac{5~{\rm TeV}}{\Lambda} \right)^2,
\label{eq:ft-sm}
\end{equation}
under the same constraint of $m_{h^0} \simlt 250~{\rm GeV}$ (our 
definitions for $m_h^2$ and $\lambda$ are $V(h) = m_h^2 |h|^2 + 
(\lambda/2) |h|^4$).  For $m_{h^0} \simlt 200~{\rm GeV}$, this number 
becomes $\approx 2.3\%$. 

Equations~(\ref{eq:ft-model},~\ref{eq:ft-sm}) imply that our theory 
achieves about a factor 4 reduction in fine-tuning.  For $\Lambda 
\approx 5~{\rm TeV}$, the scale relevant for electroweak precision 
constraints, the fine-tuning goes from ``a few percent'' to ``better 
than $10\%$'' for $m_{h^0} \simgt 200~{\rm GeV}$ (about 1~in~7 
for $m_{h^0} \simeq 250~{\rm GeV}$).  We also note that some of 
the factors included in the analysis here, for example $N_f$ in 
Eq.~(\ref{eq:f-Lambda}), are often not included in literature. 
To compare the result of our model with those of other models, 
we must take all these factors into account appropriately. 

In the next section, we extend the minimal theory presented here to 
include the $U(4)$ partners of the top quark.  This allows a further 
reduction of fine-tuning and/or a larger cutoff scale, since the top 
contribution to the divergence of the Higgs mass-squared parameter 
is then canceled by these partners.

\section{{\boldmath $U(4)$}-invariant Top Sector}
\label{sec:alternative}

In this section we extend the top quark sector of the previous 
model to include the $U(4)$ partners of the top quark.  Following 
Ref.~\cite{Chacko:2005pe}, we promote the left-handed top quark, $q$, 
and its mirror partner, $\hat{q}$, into the $U(4)$-invariant field:
\begin{eqnarray}
  Q &=& q({\bf 3},{\bf 2},1/6;\, {\bf 1},{\bf 1},0)
    + \hat{q}({\bf 1},{\bf 1},0;\, {\bf 3},{\bf 2},1/6)
\nonumber\\
  && {} 
    + q'({\bf 3},{\bf 1},2/3;\, {\bf 1},{\bf 2},-1/2)
    + \hat{q}'({\bf 1},{\bf 2},-1/2;\, {\bf 3},{\bf 1},2/3),
\label{eq:U4-LHtop}
\end{eqnarray}
where the numbers in parentheses represent gauge quantum numbers under 
$(SU(3)_A \times SU(2)_A \times U(1)_A) \times (SU(3)_B \times SU(2)_B 
\times U(1)_B)$.  The $q'$ and $\hat{q}'$ are new fields introduced 
in this procedure.  Defining the field representing the right-handed 
top quark, $\bar{u}$, and its mirror partner, $\hat{\bar{u}}$, as
\begin{equation}
  \bar{U} = \bar{u}({\bf 3}^*,{\bf 1},-2/3;\, {\bf 1},{\bf 1},0)
    + \hat{\bar{u}}({\bf 1},{\bf 1},0;\, {\bf 3}^*,{\bf 1},-2/3),
\label{eq:U4-RHtop}
\end{equation}
we can write the following $U(4)$ ($\times U(4)'$) invariant top Yukawa 
coupling:
\begin{equation}
  {\cal L}_{\rm top} 
    = y_t Q \bar{U} \Phi^\dagger,
\label{eq:U4-top-Yukawa}
\end{equation}
which contains the Yukawa couplings of Eq.~(\ref{eq:top-Yukawa}) 
when expanded in the ``component'' fields of Eqs.~(\ref{eq:U4-LHtop},%
~\ref{eq:U4-RHtop}).  The new fields $q'$ and $\hat{q}'$ in 
Eq.~(\ref{eq:U4-LHtop}) are made heavy by introducing the conjugate 
fields $q^{\prime c}({\bf 3}^*,{\bf 1},-2/3;\, {\bf 1},{\bf 2},1/2)$ and 
$\hat{q}^{\prime c}({\bf 1},{\bf 2},1/2;\, {\bf 3}^*,{\bf 1},-2/3)$ with 
the $Z_2$-invariant mass term
\begin{equation}
  {\cal L} = M(q' q^{\prime c} + \hat{q}' \hat{q}^{\prime c}).
\label{eq:Z2-mass}
\end{equation}
With the new top Yukawa coupling of Eq.~(\ref{eq:U4-top-Yukawa}), the 
only $U(4) \times U(4)'$ violating effect in the top sector is the mass 
$M$ of Eq.~(\ref{eq:Z2-mass}).  The contribution from the top quark to 
the Higgs mass-squared parameter is thus cut off at the scale $M$, which 
we take $\approx y_t f$.

The calculation of radiative corrections from the $Q$ and $\bar{U}$ 
fields to the Higgs mass-squared parameter has been performed 
in~\cite{Chacko:2005pe}.  In the present context, this translates into
\begin{equation}
  \delta m^2|_{\rm top} = -\frac{3}{8\pi^2} \frac{y_t^2 M^2}{y_t^2 f^2-M^2}
    \left( M^2\ln\frac{y_t^2 f^2+M^2}{M^2} 
    - y_t^2 f^2 \ln\frac{y_t^2 f^2+M^2}{y_t^2 f^2} \right),
\label{eq:top-contr-U4}
\end{equation}
and the top quark mass is given by
\begin{equation}
  m_t \simeq \frac{y_t M}{\sqrt{y_t^2 f^2 + M^2}}\, v\, \sin\beta.
\label{eq:U4-mtop}
\end{equation}
The top contribution of Eq.~(\ref{eq:top-contr-U4}) can be written 
in the form
\begin{equation}
  \delta m^2|_{\rm top} = -\frac{3}{8\pi^2} y_t^2 M^2\, 
    {\cal F}\biggl(\frac{y_t^2 f^2}{M^2}\biggr),
\label{eq:top-contr-U4-2}
\end{equation}
where ${\cal F}(x) \equiv \{ \ln(1+x) - x\ln(1+1/x) \}/(x-1)$ is a function 
which has the property ${\cal F}(x) = {\cal F}(1/x)$.  For $0.5 \simlt x 
\simlt 2$, this function takes values ${\cal F}(x) \simeq 0.3$.  We then 
find that, in the parameter region $0.5 \simlt y_t^2 f^2/M^2 \simlt 2$, 
the top contribution in the present model, Eq.~(\ref{eq:top-contr-U4}), 
is a factor of $(2\!\sim\!3)$ smaller than that in the previous model, 
Eq.~(\ref{eq:top-contr}), for the same value of $f$.  The contribution 
to the fine-tuning parameter from top loop, which is given by 
$\Delta^{-1}|_{\rm top} \approx m_{h^0}^2/(4 |\delta m^2|_{\rm top}| 
\sin^2\!\beta)$, is thus a factor of $(2\!\sim\!3)$ smaller than before. 
The contribution from quartic loop, $\Delta^{-1}|_{H^4}$, is the same 
and is given by Eq.~(\ref{eq:fine-tuning-Higgs}).  The fine-tuning 
parameter $\Delta^{-1}$ is given by the smaller of $\Delta^{-1}|_{\rm top}$ 
and $\Delta^{-1}|_{H^4}$, as in Eq.~(\ref{eq:fine-tuning}). 

We can now repeat the same analysis as in the previous model with 
the new $\Delta^{-1}|_{\rm top}$.  We find that under the constraint 
$m_{h^0} \simlt 250~{\rm GeV}$, the largest value of $f$ for a fixed 
$\Delta^{-1}$ is given by
\begin{equation}
  f_{\rm max} \approx 930~{\rm GeV} 
    \left( \frac{20\%}{\Delta^{-1}} \right)^{1/2},
\label{eq:fmax-U4}
\end{equation}
which occurs when $m_{h^0} \simeq 250~{\rm GeV}$, $y_t \sim \lambda \sim 2$, 
$\tan\beta \sim 1$ and $M \simeq y_t f$.  (For $M \neq y_t f$ with the 
other parameters fixed, $\Delta^{-1} \propto x/((x+1)^2 {\cal F}(x))$ where 
$x \equiv y_t^2 f^2/M^2$, so that $\Delta^{-1}$ changes only $\simlt 20\%$ 
for $0.5 \simlt y_t f/M \simlt 2$.)  For $m_{h^0} \simlt 200~{\rm GeV}$, 
this number becomes $\approx 890~{\rm GeV}$, occurring at $m_{h^0} \simeq 
200~{\rm GeV}$, $y_t \sim \lambda \sim 1.5$, $\tan\beta \simeq 1.4$ and 
$M \simeq y_t f$.  Since $\Lambda \approx 2\pi f$, we obtain the maximum 
value of the cutoff:
\begin{equation}
  \Lambda_{\rm max} \approx 6~{\rm TeV} 
    \left( \frac{20\%}{\Delta^{-1}} \right)^{1/2}.
\label{eq:Lmax-U4}
\end{equation}
For $\Delta^{-1} \simeq 10\%$, this reaches as high as $\Lambda_{\rm max} 
\approx 8~{\rm TeV}$.  In terms of the fine-tuning parameter, we find
\begin{equation}
  \Delta^{-1} \approx 28\% \left( \frac{5~{\rm TeV}}{\Lambda} \right)^2.
\label{eq:ft-model-U4}
\end{equation}
Compared with the standard model case, Eq.~(\ref{eq:ft-sm}), this is 
an improvement of a factor $\approx 8$.  This is achieved because the 
top contribution to the Higgs mass-squared parameter is cut off at the 
scale $M \approx y_t f$, without a logarithmic sensitivity to $\Lambda$.

\section{Possibility of Strong Coupling at {\boldmath $\Lambda$}}
\label{sec:strong}

In previous sections we have assumed that the theory is weakly 
coupled at $\Lambda$.  This has ensured that radiative corrections to 
the $U(4) \times U(4)'$-violating quartic terms from the gauge, Yukawa 
and $\lambda$ couplings are of order $(1/16\pi^2)\ln(\Lambda/f)$. 
If the theory is strongly coupled at $\Lambda$, i.e. $\eta \sim \eta' 
\sim 16\pi^2$, this property is not automatically guaranteed, since 
the couplings may, in general, receive corrections of order e.g. 
$\eta g^2/16\pi^2, \eta \lambda/16\pi^2 \sim 1$~\cite{Barbieri:2005ri}. 
In this section we discuss the possibility of making the theory 
strongly coupled at $\Lambda$.

Let us first consider the corrections from the coupling $\lambda$ in 
Eq.~(\ref{eq:quartic}).  We find that we can rewrite the operator in 
Eq.~(\ref{eq:quartic}) as
\begin{equation}
  {\cal L}_H = - \lambda_1 |H_A^\dagger H'_A + H_B^\dagger H'_B|^2
    - \lambda_2 |H_A^\dagger H'_A - H_B^\dagger H'_B|^2,
\label{eq:quartic-dec}
\end{equation}
where $\lambda_1 = \lambda_2 = \lambda/2$.  We then find that the first 
term preserves a $U(4)$ global symmetry under which $(H_A|H_B)$ and 
$(H'_A|H'_B)$ transform as a fundamental representation, while the 
second term preserves another $U(4)$ symmetry under which $(H_A|H_B)$ 
and $(H'_A|-\!H'_B)$ transform as a fundamental representation. 
Each of these $U(4)$'s is sufficient to protect the mass of the 
Higgs fields $h$ and $h'$, implying that the dangerous operators 
in Eq.~(\ref{eq:violating}) are generated only at order $\lambda_1 
\lambda_2 \sim \lambda^2$.  This guarantees that these operators 
receive radiative corrections only of order $16\pi^2 (\lambda/16\pi^2)^2 
\ln(\Lambda/f) \sim (1/16\pi^2)\ln(\Lambda/f)$ even at strong coupling 
(with the explicit breaking parameter $\lambda$ kept to be $O(1)$, 
of course).  We find that the collective symmetry breaking 
mechanism~\cite{Arkani-Hamed:2001nc} is automatically incorporated 
in the single operator of Eq.~(\ref{eq:quartic}) in our theory.%
\footnote{It should not be viewed that the operator in 
Eq.~(\ref{eq:quartic}) is obtained by setting the coefficients 
of the two operators in Eq.~(\ref{eq:quartic-dec}) equal by hand. 
Because of the ``chiral'' $Z_2$  symmetry, $(H_A | H_B) \leftrightarrow 
(H_A | H_B)$ and $(H'_A | H'_B) \leftrightarrow (H'_A | -\!H'_B)$, these 
coefficients are necessarily equal, $\lambda_1 = \lambda_2$, so that the 
operator of Eq.~(\ref{eq:quartic}) is really a single operator.}

We next consider the Yukawa couplings.  The Yukawa couplings, collectively 
denoted as $y$ here, connect two fermions to a scalar field $\Phi$ 
or $\Phi'$.  At order $y^2$, only the quadratic terms in the scalar 
potential receive corrections.  These terms, however, are necessarily 
$U(4) \times U(4)'$ invariant due to the $Z_2$ symmetry and the global 
$U(1) \times U(1)'$ symmetry.  The corrections to the Higgs mass-squared 
parameters, which come from the quartic terms in the scalar potential, 
thus arise only at order $y^4$.  This ensures that the dangerous operators 
receive corrections only of order $16\pi^2 (y^2/16\pi^2)^2 \ln(\Lambda/f) 
\sim (1/16\pi^2)\ln(\Lambda/f)$ even at strong coupling.

How about gauge interactions?  At the renormalizable level, we can show 
that the dangerous corrections do not arise, analogously to the case of 
the $\lambda$ coupling.  First, interactions of the form $\Phi^\dagger 
\Phi A^\mu A_\mu$ generate only the quadratic terms in the scalar 
potential, but they are always $U(4) \times U(4)'$ invariant.  The 
interactions linear in $A_\mu$ can then be decomposed into two parts, 
as in Eq.~(\ref{eq:quartic-dec}), each of which preserves a global 
$U(4)$ symmetry that is sufficient to protect the masses of $h$ and 
$h'$.  The dangerous operators thus receive corrections only of order 
$16\pi^2 (g^2/16\pi^2)^2 \ln(\Lambda/f) \sim (1/16\pi^2)\ln(\Lambda/f)$. 
This argument, however, may not apply for higher dimension operators 
suppressed by $\Lambda$, which we expect to be there.  Showing that 
the theory can really be made strongly coupled at $\Lambda$, therefore, 
requires a careful analysis of all these corrections.  Here we do not 
pursue this issue further, leaving it for future work~\cite{CGH}.

The possibility of strong coupling at $\Lambda$ is particularly 
interesting because the relation $f \approx f' \approx \Lambda/2\pi$ 
would then be naturally understood in terms of naive dimensional 
analysis~\cite{Manohar:1983md}.  It also leads to more possibilities 
for an ultraviolet theory above the scale $\Lambda$.  A closer study 
of this issue is warranted.

\section{Phenomenology}
\label{sec:pheno}

In this section we discuss the phenomenological implications of our 
models.  We mainly focus on collider signals, since the cosmological 
aspects of the models are similar to what were discussed in literature 
(see e.g.~\cite{Chacko:2005pe,Barbieri:2005ri,Berezhiani:2003xm} and 
references therein).  The only point worth mentioning is that our 
models provide a natural way to lift the mirror photon mass because 
$SU(2)_B \times U(1)_B$ is completely broken in the vacuum.  This 
relaxes many of the cosmological constraints, related to the excess 
in radiation energy density coming from the mirror sector and to the 
production of proto-Galaxies from the dark matter mirror baryons.

Regarding collider physics, our models are quite distinct.  At low 
energies the ``visible'' particle content of both models is that 
of a general two Higgs doublet standard model.  At higher energies 
additional singlet scalar particles are present.  In the model with 
the extended top sector there are new fermions charged under the 
standard model gauge group with masses of $\approx (1\!\sim\!2)~{\rm 
TeV}$, but these particles are absent in the minimal model of 
section~\ref{sec:theory}.

In general we are interested in parameter regions where no severe 
fine-tuning is required.  In these regions the lightest Higgs boson 
is relatively heavy, with masses of $O(150\!\sim\!200~{\rm GeV})$, 
allowing for an easy detection through $WW$ decays.  However, the 
detection of all five Higgs bosons at the LHC is, in general, 
non-trivial (see e.g.~\cite{Djouadi:2005gi} for recent reviews). 
Thus, without detecting the other mirror particles or singlet fields, 
our model would look simply like a two Higgs doublet standard model 
or perhaps even just the standard model.

It is important to consider whether one can have additional signals 
at the LHC that allow to distinguish the models from a simple two Higgs 
doublet standard model.  In the model of section~\ref{sec:alternative}, 
there are colored fermions of masses $\approx (1\!\sim\!2)~{\rm 
TeV}$, which can be found easily at the LHC.  In the model of 
section~\ref{sec:theory}, however, the detection of new physics 
beyond the two Higgs doublet standard model will, at best, be 
a difficult task, because all the new particles are singlet under 
the standard model gauge group. 

The simplest possibility would be to look for invisible decays of the 
Higgs bosons into mirror fermions~\cite{Foot:1991bp}.  The relevant 
vertices, however, arise only from the mixing between the neutral 
scalars of the two sectors, so that they are all suppressed by 
powers of $v/f$ or $v/f'$.  The most important decay channels 
would be to the mirror bottom quark, which is the heaviest mirror 
particle available below the Higgs boson masses.  The branching 
ratios to these invisible decay modes, however, are still too small 
to be observed at the LHC, which requires the product of the Higgs 
production cross section normalized to the standard model one and 
the branching fraction into the invisible channel to be about $0.1$ 
or larger~\cite{ATLAS-inv}.

The situation is similar in pseudoscalar and charged Higgs boson decays. 
For the pseudoscalar case, the vector boson fusion channel is not 
available because it does not couple to the gauge bosons at tree level. 
This makes it almost impossible to detect the invisible width because 
of the standard model background.  For the charged Higgs boson case, one 
might try to tag the associated production of a visible charged particle 
and invisible fields.  The decay modes of the charged Higgs boson into 
something visible and mirror particles, however, proceed only through 
higher dimension operators and are highly suppressed.  The other 
possibility would be to look at the cascade of a charged Higgs into 
mirror particles through a neutral Higgs, for example as $H^+ \rightarrow 
W h^0 \rightarrow l + {\rm missing\: energy}$, where $H^+$ is produced 
in the standard way through $gb \rightarrow t H^+$~\cite{ATLAS-gb}. 
This requires, however, to fully reconstruct the top quark hadronically, 
and a large standard model background from $gg \rightarrow \bar{t} t$ 
with one top quark decaying semileptonically makes the observation of 
these ``semi-invisible'' decay impractical~\cite{Bisset:2000ud}. 

A possibility of distinguishing our model from a simple two Higgs 
doublet standard model may come from the decay of the radial excitation 
modes of $\Phi$ and $\Phi'$, which have the masses $\sqrt{2\eta} f$ 
and $\sqrt{2\eta'} f'$.  Since the light Higgs boson has mixings with 
these modes of $O(v/f)$, a heavy radial mode can be produced instead of 
a Higgs boson~\cite{Schabinger:2005ei} with a production cross section 
a factor of $\simeq v^2/f^2$ below that of the Higgs boson.  The radial 
modes have couplings of $O(f)$ to a pair of the light Higgs bosons. 
Thus, after being produced on-shell, the radial mode can decay into 
two light Higgs bosons, which then decay into standard model particles. 
This may be the dominant decay of the radial modes; the only competing 
ones would be decays into a mirror top or gauge boson pair, which could 
be kinematically forbidden because all of these particles have masses 
of $O(f)$.  We expect that the masses of the $CP$ even Higgs bosons 
are above the $ZZ$ threshold, so one can look at the light Higgs bosons, 
which are produced by the decay of a radial mode and decay into $ZZ$ and 
$WW$ pairs.  A rough estimate, however, shows that the channel in which 
one $Z$ and one $W$ decay leptonically does not have a large event rate. 
Thus, even though the standard model background is small, it is difficult 
to observe this channel (unless the mass of the radial mode is somewhat 
unexpectedly small).  One may also look for (one of) the Higgs bosons 
decaying either into $b$ or $\tau$ pairs, which increases the event rate. 
This, however, also increases the standard model background, so that 
a more detailed analysis is needed to see if this mode is useful. 
We also note that if $\eta$ and $\eta'$ are of order unity or somewhat 
larger, the radial modes become (much) heavier than a TeV, and the 
detection of these modes at the LHC becomes almost impossible. 
Following the discussion after Eq.~(\ref{eq:f-Lambda}), this may 
be the case for smaller fine-tuning.

At a linear collider, invisible decays of the Higgs bosons may be 
accessible.  The branching ratios, however, are not so large $\approx 
10^{-3}$ for the associated production with a $Z$ boson, $ZH \rightarrow 
ll + {\rm missing\: energy}$, so that it is not clear if this can 
be detected.  Another possible channel is to produce a pseudoscalar 
associated with a neutral Higgs boson, and look for an invisible decay 
of the pseudoscalar Higgs boson, which has a branching ratio of order 
$(v/f)^2$.  The precise study of the masses and couplings of the Higgs 
bosons may also be used to discriminate between our model and the 
simple two Higgs doublet standard model.

Finally, precision electroweak constraints are easily satisfied 
by construction~\cite{Chacko:2005pe}.  In the model of 
section~\ref{sec:theory} all particles beyond those in the two 
Higgs doublet standard model are singlet under the standard model 
gauge group, and the effects on the precision electroweak observables 
are small.  The contributions from heavy radial modes, for example, come 
only through mixings with light Higgs bosons, which induce an additional 
logarithmic contribution which effectively appears as arising from two 
very heavy Higgs bosons but with the coefficients suppressed by factors 
of ${\cal O}(v^2/f^2)$.  The contributions from vector-like fermions 
in the model of section~\ref{sec:alternative} are also small, 
as they have $SU(2)_L \times U(1)_Y$ invariant masses of order 
$(1\!\sim\!2)~{\rm TeV}$.  Dangerous operators induced by ultraviolet 
physics are either suppressed by assumed symmetries or, if it is not 
possible, as in the case of the operator for the $S$ parameter, are 
suppressed by our rather high cutoff scale of $(5\!\sim\!8)~{\rm TeV}$.

\section{Conclusions}
\label{sec:concl}

In this paper we have constructed a theory in which radiative corrections 
to the quadratic part of the potential are constrained to be symmetric 
under a global $U(4) \times U(4)'$ symmetry due to a discrete $Z_2$ 
symmetry, while the quartic part does not possess this symmetry at all. 
The theory is weakly coupled at the cutoff scale $\Lambda$, and has 
a simple structure where the two Higgs doublet standard model is 
simply ``twinned'' due to the $Z_2$ symmetry.  The two Higgs doublets 
have a quartic coupling at tree level, while their squared masses are 
generated only at order $f^2/16\pi^2$, where $f$ is an order parameter 
for the $U(4) \times U(4)'$ breaking, which is supposed to be a factor 
of $4\pi$ smaller than $\Lambda$.  This setup, thus, potentially 
allows us to have a large hierarchy between the electroweak VEV, $v$, 
and the cutoff scale.

We have carefully studied fine-tuning in this theory and found that 
we do not obtain a hierarchy as large as what may naively be expected, 
$\Lambda \simeq 4\pi f \simeq 16\pi^2 v$.  The theory, however, still 
allows a reduction of fine-tuning by a factor of $\approx 4$ compared 
with the standard model, even in the minimal version, which allows us 
to push up the cutoff scale to about $5~{\rm TeV}$ without significant 
fine-tuning ($\Delta^{-1} \approx 14\%$).  This is almost enough to solve 
the little hierarchy problem, implied by the mismatch between the stability 
of the electroweak scale and the constraints from experiments.  With the 
$U(4)$-extended top quark sector, we can further reduce fine-tuning to 
the level of $30\%$ for $\Lambda \approx 5~{\rm TeV}$, or if we allows 
a mild tuning of order $10\%$ the cutoff scale can be as high as 
$\Lambda \approx 8~{\rm TeV}$.  In general, the theory prefers a heavy 
Higgs boson, $m_{h^0} \simeq (150\!\sim\!250)~{\rm GeV}$, and small 
values for the ratio of the VEVs for the two Higgs fields, $\tan\beta 
\simeq (1\!\sim\!2)$. 

Our theory provides an example of a potentially embarrassing situation 
at the LHC.  While the theory is not significantly fine-tuned, the LHC 
may just see the two Higgs doublet standard model, and may not find 
any new physics responsible for cutting off the divergences of the 
Higgs mass-squared parameter.  This occurs in the model without extra 
vector-like fermions.  All quadratic divergences in the Higgs mass-squared 
parameter due to standard model loops are canceled by fields that are 
singlet under the standard model gauge group.  The deviations from 
the simple two Higgs doublet model due to these singlet fields can be 
very small at the LHC.  We have discussed several possible processes 
that may be able to discriminate our model from the two Higgs doublet 
model at the LHC and at a linear collider.  It will be interesting 
to study these processes in more detail.

Possible physics above the cutoff scale $\Lambda$ is unknown.  We have 
discussed the possibility of extending the theory to the strongly coupled 
regime at $\Lambda$.  It would be interesting to pursue possible 
ultraviolet physics that reduces to our theory below the scale of 
$\Lambda \approx (5\!\sim\!8)~{\rm TeV}$.

\vspace{0.4cm}

{\bf Note added:}

While completing this paper, we received Ref.~\cite{BH}, which also 
addresses the little hierarchy problem in the context of the two Higgs 
doublet standard model.

\section*{Acknowledgments}

The work of Z.C. was supported by the National Science Foundation 
under grant PHY-0408954.  The work of Y.N., M.P. and G.P. was supported 
in part by the Director, Office of Science, Office of High Energy 
and Nuclear Physics, of the US Department of Energy under Contract 
DE-AC02-05CH11231 and by the National Science Foundation under grant 
PHY-00-98840.  The work of Y.N. was also supported by the National 
Science Foundation under grant PHY-0403380, by a DOE Outstanding Junior 
Investigator award, and by an Alfred P. Sloan Research Fellowship.

\end{document}